\documentclass[a4paper]{article}
\usepackage{pdfsync}
\usepackage[latin1]{inputenc}
\usepackage{graphicx}
\usepackage{amssymb,amsfonts,amsmath}

\usepackage{theorem}
\usepackage{hyperref}
\usepackage{graphicx}
\usepackage{multirow}
\usepackage{verbatim}

\usepackage{color}

\newtheorem{theorem}{Theorem}[section]
\newtheorem{corollary}{Corollary}[section]
\newtheorem{lemma}{Lemma}[section]
\newtheorem{proposition}{Proposition}[section]
{\theorembodyfont{\rmfamily}

\newtheorem{remark}{Remark}[section]
\newtheorem{example}{Example}[section]
}
\newenvironment{proof}[1][Proof]{\noindent\textbf{#1.} }{\newline \hspace*{\textwidth}\hspace*{-0,4cm} \rule{0.5em}{0.5em} \vspace{0,2cm}}

\begin{document}

\title{\textbf{Relative velocities, geometry, and expansion of space}}

\author{Vicente J. Bol\'os \\{\small Dpto. Matem\'aticas para la Economía y la Empresa, Facultad de Econom\'{\i}a,}\\ {\small Universidad de Valencia. Avda. Tarongers s/n. 46022, Valencia,
Spain.}\\ {\small e-mail\textup{: \texttt{vbolos@uv.es}}}\\\\
Sam Havens\\{\small Department of Mathematics and Interdisciplinary Research Institute for the Sciences,}\\ {\small California State University, Northridge, USA}\\ {\small e-mail\textup{: \texttt{samhavens@gmail.com}}}\\\\
David Klein\\{\small Department of Mathematics and Interdisciplinary Research Institute for the Sciences,}\\ {\small California State University, Northridge, USA}\\ {\small e-mail\textup{: \texttt{david.klein@csun.edu}}}}
\date{\today}

\maketitle

\begin{abstract}
What does it mean to say that space expands? One approach to this question is the study of relative velocities.  In this context, a non local test particle is ``superluminal'' if its relative velocity exceeds the local speed of light of the observer.  The existence of superluminal relative velocities of receding test particles, in a particular cosmological model, suggests itself as a possible criterion for expansion of space in that model.    In this point of view, superluminal velocities of distant receding galaxy clusters result from the expansion of space between the observer and the clusters. However, there is a fundamental ambiguity that must be resolved before this approach can be meaningful.  The notion of relative velocity of a nonlocal object depends on the choice of coordinates, and this ambiguity suggests the need for coordinate independent definitions.  In this work, we review four (inequivalent) geometrically defined and universal notions of relative velocity: Fermi, kinematic, astrometric, and spectroscopic relative velocities.   We apply this formalism to test particles undergoing radial motion relative to comoving observers in expanding Robertson-Walker cosmologies, and include previously unpublished results on Fermi coordinates for a class of inflationary cosmologies.  We compare relative velocities to each other, and show how pairs of them determine geometric properties of the spacetime, including the scale factor with sufficient data.  Necessary and sufficient conditions are given for the existence of superluminal recessional Fermi speeds in general Robertson-Walker cosmologies.  We conclude with a discussion of expansion of space.
\end{abstract}

\vspace{10 pt}


\section{Introduction}

General relativity restricts the speed of a test particle to be less than the speed of light relative to an observer at the exact spacetime point of the test particle, but for test particles and observers located at different space-time points, the theory provides no \textit{a priori} definition of relative velocity.  Different coordinate charts give rise to different relative velocities.

This is perhaps most convincingly illustrated by the Milne universe, characterized by scale factor $a(t)=t$, where $t$ is cosmological time (for the metric see \eqref{eqds2} below, with $k=0$).  According to Hubble's law, whose formulation uses standard curvature coordinates,

\begin{equation*}
\label{introhubble}
\dot{d}(t)\equiv v_{H}=Hd,
\end{equation*}
where $H\equiv\dot{a}(t)/a(t)$ is the Hubble parameter, $d$ is the proper distance at fixed time $t$ from the observer to a comoving test particle, and the overdot signifies differentiation with respect to $t$.  For the Milne universe, $H=1/t >0$, and thus at any time $t$ and at sufficiently large proper distance $d$, the relative speed, $v_{H}$, of the test particle necessarily exceeds the local speed of light for the observer. However, with a simple change of coordinates, $\tau=t\cosh\chi$ and $\rho=t\sinh\chi$,
the metric \eqref{eqds2} is transformed to the Minkowski metric, and the Milne universe becomes the forward light cone of Minkowski spacetime.  In these  coordinates there are no superluminal speeds.

The ambiguity illustrated by this example has analogs in all spacetimes, and this feature led to consideration of the need for a strict definition of ``radial
velocity'' within the solar system at the General Assembly of
the International Astronomical Union (IAU), held in 2000 (see
\cite{Soff03,Lind03}).

Thereafter, a series of papers \cite{Bolos02,Bolos05,Bolos07} appeared addressing the general question of relative velocities and culminated in the introduction of four geometrically defined (but inequivalent) notions of relative velocity: Fermi, kinematic, astrometric, and the spectroscopic relative velocities. All four relative velocities have physical justifications and have been used to study properties of spacetimes (see \cite{KC10,Klein11,Bolos12}).

Two distinct notions of simultaneity play roles in the four definitions of relative velocities: ``spacelike simultaneity'' (also described as ``Fermi simultaneity'', see \cite{Ferm22}) and ``lightlike simultaneity.''
The Fermi and kinematic relative velocities are defined in terms of spacelike simultaneity, according to which events are simultaneous if they lie on the same Fermi space slice determined by a fixed Fermi time coordinate. For a test particle undergoing radial motion, the Fermi relative velocity, $v_{\mathrm{Fermi}}$ is the rate of change of proper distance of the test particle away from the central observer along the Fermi space slice with respect to proper time of the observer.
The kinematic relative velocity is found by first parallel transporting the 4-velocity $u'$ of the test particle at the spacetime point $q_{\mathrm{s}}$, along a radial spacelike geodesic (lying on a Fermi space slice) to a 4-velocity denoted by $\tau_{q_{\mathrm{s}}p}u'$ in the tangent space of the observer at spacetime point $p$, whose 4-velocity is $u$. The kinematic relative velocity $v_{\mathrm{kin}}$ is then the unique vector orthogonal to $u$, in the tangent space of the observer, satisfying $\tau_{q_{\mathrm{s}}p}u'=k (u+v_{\mathrm{kin}})$ for some scalar $k$ (which is easily shown to be uniquely determined).

The spectroscopic (or barycentric) and astrometric relative velocities can be found, in principle, from spectroscopic and astronomical observations.  Mathematically, both rely on the notion of ``lightlike simultaneity'', according to which two events are simultaneous if they both lie past-pointing horismos (which is tangent to the backward light cone) at the spacetime point $p$ of the central observer. The spectroscopic relative velocity $v_{\mathrm{spec}}$ is calculated analogously to $v_{\mathrm{kin}}$, described in the preceding paragraph, except that the 4-velocity $u'$ of the test particle is parallel transported to the tangent space of the observer along a null geodesic lying on the past-pointing horismos of the observer, instead of along the Fermi space slice.  The astrometric relative velocity, $v_{\mathrm{ast}}$, of a test particle whose motion is purely radial is calculated analogously to $v_{\mathrm{Fermi}}$, as the rate of change of the affine distance, which corresponds to the \textit{observed} proper distance (through light signals at the time of observation) with respect to the proper time of the observer, as may be done via parallax measurements. We describe this more precisely in the sequel, and complete definitions for arbitrary (not necessarily radial) motion may be found in \cite{Bolos07}.

In this work, we review and explain these four notions of relative velocities in the context of test particles receding radially from comoving observers in Robertson-Walker cosmologies.  This particular scenario lends itself to a consideration of a possible meaning for the expansion of space.  The existence of a superluminal relative velocity of receding test particles, in a particular cosmological model, is a possible criterion for expansion of space in that model\footnote{One must, of course, take into account an ambiguity.  Different space slices are associated with different coordinate systems.}. In this framework, superluminal velocities of distant receding galaxy clusters, in the actual universe, are the result of the expansion of space between the observer and the clusters.

Let us make precise the concept ``superluminal''. Let $v(q)$ be a relative velocity of a test particle at an event $q$ with respect to an observer at an event $p$, and let $c(q)$ be the corresponding relative velocity of a photon at the same event $q$ (note that any well-posed concept of relative velocity can be extended to photons) with the same spatial direction as the particle and with respect to the same observer at $p$. Then $\|v(q)\|<\|c(q)\|$ always, but $\|v(q)\|$ can exceed the local speed of light at $p$, which we take as $c=1$ throughout the paper. In this case, we say that the particle is ``superluminal'', but that does not mean that it travels faster than light. In the particular case of the relative velocities introduced above, $\|c(q)\|$ is always $1$ for the kinematic and spectroscopic velocities, and so, there do not exist superluminal spectroscopic or kinematic velocities. On the other hand, the Fermi and astrometric relative velocities of a photon can be less than $1$, equal to $1$, or greater than $1$, allowing for the possibility superluminal velocities.

Much of the material we present here is based on references \cite{Klein11} and \cite{BolosKlein}. Exact Fermi coordinates were found in \cite{Klein11} for expanding Robertson-Walker spacetimes and Fermi charts were shown to be global for non inflationary scale factors\footnote{A scale factor $a(t)$ is non inflationary if $\ddot{a}(t)\leq0$ for all $t$.}. These coordinates were then used to calculate the (finite) diameter of the Fermi space slice, as a function of the observer's proper time, and  Fermi velocities of (receding) comoving test particles.  Reference \cite{BolosKlein} extended the results of \cite{Klein11}  by finding general formulas for all four relative velocities for test particles undergoing arbitrary radial motion in expanding Robertson-Walker spacetimes, finding relationships among these relative velocities, and showing that their ratios determine geometric properties of the spacetime.  Those results were illustrated in the de Sitter universe, the radiation-dominated universe, the matter-dominated universe, and more generally, for cosmologies for which the scale factor, $a(t)=t^{\alpha}$ for $0< \alpha <1$.

Previously unpublished results \cite{Havens} are also included in Section \ref{fermi} of this work, which provides a proof that  Fermi coordinates extend to the cosmological event horizon in inflationary cosmologies with scale factors of the form $a(t)=t^{\alpha}$ for $\alpha >1$.  This allows for interesting comparisons between inflationary and non inflationary expanding cosmologies, which we discuss in the final section.

\section{The Robertson-Walker metric}
\label{sec:2}

The Robertson-Walker metric in curvature-normalized coordinates (or Robertson-Walker coordinates) is given by the line element
\begin{equation}
\label{eqds2}
\mathrm{d}s^2=-\mathrm{d}t^2+a^2(t)\left( \mathrm{d}\chi ^2 +S_k^2\left( \chi \right) \mathrm{d}\Omega ^2 \right) ,
\end{equation}
where $\mathrm{d}\Omega =\mathrm{d}\theta +\sin ^2\theta \mathrm{d}\varphi $, $a(t)$ is a positive and increasing scale factor, with $t>0$, and
$ S_k\left( \chi \right) = \sin \left( \chi \right) $ if $k=1$, $\chi$ if $k=0$, or $\sinh \left( \chi \right) $ if $k=-1$.
There is a coordinate singularity in \eqref{eqds2} at $\chi =0$, but this will not affect the calculations that follow. Since our purpose is to study radial motion with respect to a central observer, it suffices to consider the $2$-dimensional Robertson-Walker metric given by
\begin{equation}
\label{metric}
\mathrm{d}s^2=-\mathrm{d}t^2+a^2(t)\mathrm{d}\chi ^2,
\end{equation}
for which there is no singularity at $\chi=0$.  We assume throughout that $a(t)$ is a smooth, increasing function of $t>0$.

\section{Notation}
\label{notation}

We denote a central observer located at $\chi=0$ by the world line, $\beta (\tau )=(\tau ,0)$, and a test particle by $\beta ' \left( \tau '\right) =\left( t\left( \tau '\right) , \chi \left( \tau '\right) \right) $. Each spacetime path is parameterized by its own proper time, and we will always assume that $\chi >0$ for the latter.
We have to remark that in the case $k=1$, the coordinate $\chi$ is upper bounded by $\pi $; thus, for simplicity, we assume throughout that $k=-1$ or $0$, but our methods also work for the case $k=1$ by restricting $\chi$ to $\left] 0, \pi \right[ $.

Our aim is to study the relative velocities of $\beta ' $ with respect to, and observed by, $\beta $.
Denote the 4-velocity of $\beta $ by $U$ and identify $U=\frac{\partial }{\partial t}=(1,0)$. Similarly, denote the 4-velocity of $\beta ' $ by $U'=\dot{t}\frac{\partial }{\partial t}+\dot{\chi }\frac{\partial }{\partial \chi }=( \dot{t},\dot{\chi })$, where the overdot indicates differentiation with respect to $\tau '$, the proper time of $\beta ' $. From $g(U',U')=-1$, we find,
\begin{equation}
\label{dott}
\dot{t}=\sqrt{a^2(t)\dot{\chi }^2+1}.
\end{equation}

Vector fields will be represented by upper case letters, and vectors (in the tangent space at a single spacetime point) by lower case letters. Following this notation, the 4-velocity of $\beta $ at a fixed event $p=(\tau ,0)$ will be denoted by $u=(1,0)$. The Fermi, kinematic, spectroscopic, and astrometric relative velocity vector fields (to be defined below) are denoted respectively as $V_{\mathrm{kin}}$, $V_{\mathrm{Fermi}}$, $V_{\mathrm{spec}}$ and $V_{\mathrm{ast}}$.  We shall see that they are vector fields defined on the spacetime path, $\beta $, i.e., the central observer.  By definition,  all of these relative velocities are spacelike and orthogonal to $U$, so that they are each proportional to the unit vector field $\mathcal{S}:=\frac{1}{a(t)}\frac{\partial }{\partial \chi }$\footnote{This should not to be confused with the relative position vector field $S$ used in \cite{Bolos07}; in fact, $\mathcal{S}$ is the normalized version of $S$.}.

\begin{figure}[tbp]
\begin{center}
\includegraphics[width=0.35\textwidth]{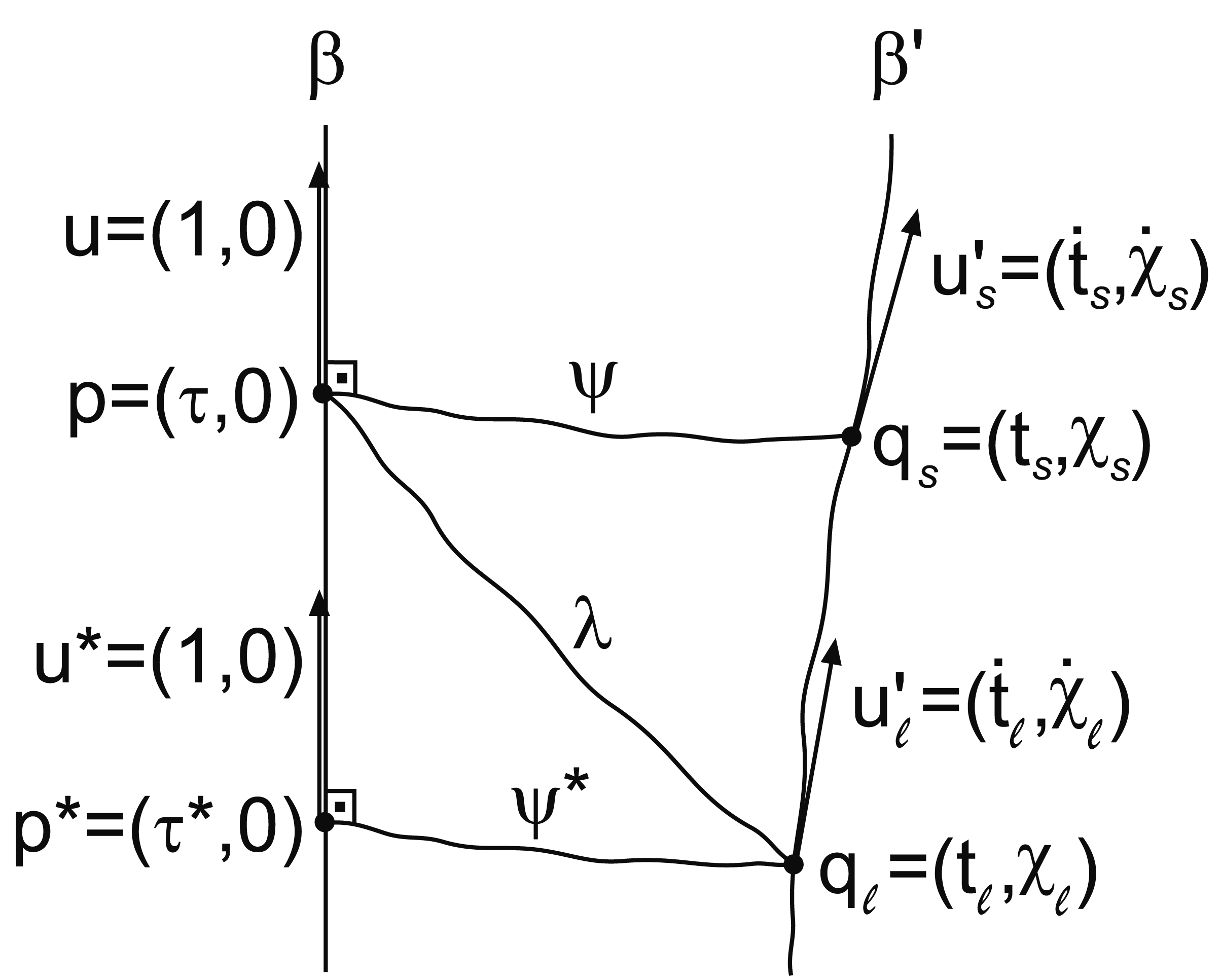}
\end{center}
\caption{Scheme of the elements involved in the study of the relative velocities of a test particle $\beta ' $ with respect to the central observer $\beta $. The curves $\psi $ and $\psi ^*$ are spacelike geodesics orthogonal to the 4-velocity of $\beta $, and $\lambda $ is a lightlike geodesic.}
\label{diagram}
\end{figure}

It is possible to compare the four relative velocities for a given test particle.  Direct comparisons may be made of the Fermi and kinematic relative velocities, because of the common dependence of these two notions of relative velocity on spacelike simultaneity.  Similarly, direct comparisons of the astrometric and spectroscopic relative velocities are also possible.  However, a comparison of all four relative velocities made at a particular instant by the central observer $\beta $ is possible only with data from two different spacetime events ($q_{\mathrm{s}}$ and $q_{\ell}$ in Figure \ref{diagram}) of the test particle.  Such a comparison, to which we refer as an \textit{instant comparison}, therefore lacks physical significance, unless the evolution of the test particle $\beta ' $ can be deduced from its 4-velocity at one spacetime point, e.g. for comoving or, more generally, geodesic test particles.

It will also be possible to compare all four relative velocities at a fixed spacetime event $q_{\ell}$ of the test particle through observations from two different times of the central observer, identified as $\tau$ and $\tau^*$ in Figure \ref{diagram}. In the sequel, we refer to such a comparison of the relative velocities as a \textit{retarded comparison}.

In all that follows, we use the following notation for vectors at a given spacetime point $p=(\tau,0)$ and a spacetime point $p^*=(\tau^*,0)$ in the past of $p$ (see Figure \ref{diagram}): $v_{\mathrm{kin}}=V_{\mathrm{kin}~p}$, $v_{\mathrm{Fermi}}=V_{\mathrm{Fermi}~p}$, $v_{\mathrm{kin}}^*=V_{\mathrm{kin}~p^*}$, $v_{\mathrm{Fermi}}^*=V_{\mathrm{Fermi}~p^*}$, $v_{\mathrm{spec}}=V_{\mathrm{spec}~p}$ and $v_{\mathrm{ast}}=V_{\mathrm{ast}~p}$.
So, in an instant comparison we compare $v_{\mathrm{kin}}$, $v_{\mathrm{Fermi}}$, $v_{\mathrm{spec}}$, and $v_{\mathrm{ast}}$, while in a retarded comparison we compare $v_{\mathrm{kin}}^*$, $v_{\mathrm{Fermi}}^*$, $v_{\mathrm{spec}}$, and $v_{\mathrm{ast}}$.

\section{Spacelike simultaneity and Fermi coordinates}
\label{seckinf1}

Let $p=(\tau ,0)$ be an event of the central observer $\beta $ with $4$-velocity $u$. An event $q$ is \textit{spacelike} (or \textit{Fermi}) \textit{simultaneous} with $p$ if $g(\exp_p^{-1}q,u)=0$\footnote{The exponential map, $\exp_{p}v$, denotes the evaluation at affine parameter $1$ of the geodesic starting at the point $p$, with initial derivative $v$.}. The \textit{Fermi space slice} $\mathcal{M}_{\tau}$ consists of all the events that are spacelike simultaneous with $p$\footnote{See \cite{Klein11}; $\mathcal{M}_{\tau}$ is also called the \textit{Landau submanifold} and denoted by $L_{p,u}$ in \cite{Bolos02}.}.

More explicitly, let $\rho$ denote proper length along a spacelike geodesic orthogonal to $\beta $.  Then, the vector field,
\begin{equation}
\label{eqX}
X:=\frac{\partial}{\partial \rho}=\frac{d t}{d \rho}\frac{\partial }{\partial t}+\frac{d \chi}{d \rho}\frac{\partial }{\partial \chi}=-\sqrt{\left( \frac{a(\tau )}{a(t)}\right) ^2-1}\frac{\partial }{\partial t}+\frac{a(\tau )}{a^2(t)}\frac{\partial }{\partial \chi},
\end{equation}
is geodesic, spacelike, unit, and $X_p$ is orthogonal to the 4-velocity $u=(1,0)$ at $p$; i.e., $X_p$ is tangent to $\mathcal{M}_{\tau }$. Let $q_{\mathrm{s}}=\left( t_{\mathrm{s}},\chi _{\mathrm{s}} \right) $ be the unique event of $\beta ' \cap \mathcal{M}_{\tau }$. Then there exists an integral curve of $X$ from $p$ to $q_{\mathrm{s}}$ (the geodesic $\psi $ in Figure \ref{diagram}), and so, using  \eqref{eqX} we can find a relationship between $\tau $, $t_{\mathrm{s}}$ and $\chi _{\mathrm{s}}$:
\begin{equation}
\label{eqt1}
\int _{\tau }^{t_{\mathrm{s}}} \frac{a(\tau )}{a^2(t)}\frac{-1}{\sqrt{\left( \frac{a(\tau )}{a(t)}\right) ^2-1}}\,\mathrm{d}t =\int _0^{\chi _{\mathrm{s}}} \,\mathrm{d}\chi
\quad \Longrightarrow \quad
\int _{t_{\mathrm{s}}}^{\tau } \frac{a(\tau )}{a(t)}\frac{1}{\sqrt{a^2(\tau )-a^2(t)}}\,\mathrm{d}t =\chi _{\mathrm{s}}.
\end{equation}
From \eqref{eqt1} we see that $t_{\mathrm{s}}<\tau $, and more generally, it follows from \eqref{eqX} that the time coordinate $t$ is a decreasing function of proper length $\rho$ along any spacelike geodesic orthogonal to the central observer's worldline.
Moreover, since we only consider positive coordinate times, we have to impose $t_{\mathrm{s}}>0$ and then it is necessary that $0<\chi _{\mathrm{s}}<\chi _{\mathrm{smax}}$, where
\begin{equation}
\label{eqchiSmax}
\chi _{\mathrm{smax}}:=\int _{0}^{\tau } \frac{a(\tau )}{a(t)}\frac{1}{\sqrt{a^2(\tau )-a^2(t)}}\,\mathrm{d}{t}.
\end{equation}

The metric \eqref{eqds2} may be expressed in the Fermi coordinates of the central observer $\beta $ (or Fermi observer in this context).  By design of Fermi coordinates, $\tau = t$ on the path $\beta (t)$ of the Fermi observer (where $\rho =0$), but the two time coordinates differ away from that path. For expanding Robertson-Walker spacetimes, it was shown in \cite{Klein11} that Fermi coordinates are global when the spacetime is non inflationary, and we show in Section \ref{fermi} that Fermi coordinates extend to the cosmological event horizon when the scale factor takes the form $a(t)=t^{\alpha}$ for $\alpha >1$.

In any case, for two spacetime dimensions, it may be shown that the metric \eqref{metric} expressed in Fermi coordinates $(\tau, \rho)$ takes the form,
\begin{equation}\label{fermipolar}
\mathrm{d}s^2=g_{\tau\tau}(\tau, \rho) \mathrm{d}\tau^2+\mathrm{d}\rho^2.
\end{equation}
General formulas for $g_{\tau \tau }$ are provided in \cite{Klein11, BolosKlein}, but we include in Section \ref{sec:2.2} a self-contained description of $g_{\tau \tau }$ in terms of the Fermi and kinematic relative velocities of test particles, relative to the central observer.

Setting $\mathrm{d}s^2=0$ in \eqref{fermipolar}, shows that the velocity  of a distant radial photon with spacetime coordinates $(\tau,\rho)$, relative to the Fermi observer $\beta $, is given by $|d\rho/d\tau|= \sqrt{-g_{\tau\tau}(\tau,\rho)}$.  Thus, the metric \eqref{fermipolar} may be understood as a natural generalization of the Minkowski metric when the speed of a photon depends on its spacetime coordinates, and may expressed as $\mathrm{d}s^2=-c(\tau,\rho)^2 \mathrm{d}\tau^2+\mathrm{d}\rho^2$ where $c(\tau,\rho)$ is the Fermi speed of a photon at the spacetime point $(\tau,\rho)$ relative to the Fermi observer located at $(\tau,0)$.

\section{Fermi and kinematic relative velocities}
\label{sec:2.2}

Based on the previous section, we may write the 4-velocity of the radially moving test particle,  $\beta '$, using Fermi coordinates as,  $u'_{\mathrm{s}}=\left. \dot{\tau}\frac{\partial }{\partial \tau }\right| _{q_{\mathrm{s}}} +\left. \dot{\rho}\frac{\partial }{\partial \rho }\right| _{q_{\mathrm{s}}}=(\dot{\tau},\dot{\rho})$, where the overdot signifies differentiation with respect to proper time of $\beta '$.

Let $p=(\tau ,0)$ be an event of $\beta $ from which we measure relative velocities. The Fermi relative velocity of $u'_{\mathrm{s}}$ with respect to $u=(1,0)$ is given by,
\begin{equation}
\label{fermidefinition2}
v_{\mathrm{Fermi}}=\frac{d\rho}{d\tau}\mathcal{S}_p=\frac{\dot{\rho}}{\dot{\tau}}\mathcal{S}_p,
\end{equation}
where $\mathcal{S}_p= \partial/\partial\rho|_p$.  For computing $v_{\mathrm{Fermi}}$, observe first from \eqref{eqX}, that,
\begin{equation}
\label{eqrho}
\rho=\rho (t_{\mathrm{s}},\chi _{\mathrm{s}})=\int _{t_{\mathrm{s}}}^{\tau (t_{\mathrm{s}},\chi _{\mathrm{s}})}\frac{a(t)}{\sqrt{a^2(\tau (t_{\mathrm{s}},\chi _{\mathrm{s}}))-a^2(t)}}\,\mathrm{d}t,
\end{equation}
where the function $\tau (t_{\mathrm{s}},\chi _{\mathrm{s}})$ is defined implicitly by \eqref{eqt1}. Then, applying \eqref{eqrho} in \eqref{fermidefinition2} we have,
\begin{equation*}
\label{vfermigeneral}
v_{\mathrm{Fermi}} =\frac{\dot{\rho }}{\dot{\tau }}\mathcal{S}_p=\frac{\frac{\partial \rho }{\partial t_{\mathrm{s}}}\dot{t}_{\mathrm{s}}+\frac{\partial \rho }{\partial \chi _{\mathrm{s}}}\dot{\chi }_{\mathrm{s}}}{\frac{\partial \tau }{\partial t_{\mathrm{s}}}\dot{t}_{\mathrm{s}}+\frac{\partial \tau }{\partial \chi _{\mathrm{s}}}\dot{\chi }_{\mathrm{s}}}\mathcal{S}_p.
\end{equation*}
In order to define kinematic relative velocity, we need some additional notation. Let $\tau _{q_{\mathrm{s}} p}$ represent the parallel transport from $q_{\mathrm{s}}$ to $p$ by the unique geodesic joining $q_{\mathrm{s}}$ and $p$, named $\psi $ in Figure \ref{diagram}. Following the description provided in the introduction, the kinematic relative velocity $v_{\text{kin}}$ of $u'$ with respect to the central observer's four-velocity $u$ is given by (see \cite{Bolos07}),
\begin{equation}
\label{vk1}
v_{\text{kin}}=\frac{1}{-g\left( \tau _{q_{\mathrm{s}} p}u'_{\mathrm{s}},u\right)}\tau _{q_{\mathrm{s}} p}u'-u.
\end{equation}
It is a general property of the kinematic relative velocity, that its magnitude, $\|v_{\mathrm{kin}}\|<1$. Using the relations $g\left( \tau _{q_{\mathrm{s}} p}u'_{\mathrm{s}},\partial/\partial\rho|_p\right) =g\left( u'_{\mathrm{s}},\partial/\partial\rho|_{q_{\mathrm{s}}}\right) $ and $g\left( \tau _{q_{\mathrm{s}} p}u'_{\mathrm{s}},\tau _{q_{\mathrm{s}} p}u'_{\mathrm{s}}\right) =-1$, , we can find $\tau _{q_{\mathrm{s}} p}u'_{\mathrm{s}}$, and then obtain the kinematic relative velocity of $u'_{\mathrm{s}}$ with respect to $u$ as,
\begin{equation}
\label{vkinfermi}
v_{\mathrm{kin}}=\frac{1}{\sqrt{-g_{\tau\tau}(\tau,\rho)}}\frac{d\rho}{d\tau}\mathcal{S}_p.
\end{equation}
Comparing \eqref{fermidefinition2} and \eqref{vkinfermi}, we see that the kinematic and Fermi relative velocities of an arbitrary test particle at the spacetime point $(\tau,\rho)$ determine $g_{\tau\tau}(\tau,\rho)$. We state this in the form of a proposition:

\begin{proposition}\label{findmetricA}
For a Robertson-Walker spacetime with scale factor $a(t)$ that is a smooth, increasing, unbounded function of $t$, the kinematic and Fermi speeds of any test particle undergoing radial motion with respect to a comoving observer determine the Fermi metric tensor element $g_{\tau\tau}$ at the spacetime point of the particle, via,
\begin{equation*}
g_{\tau\tau}(\tau,\rho)=-\frac{\|v_{\mathrm{Fermi}}\|^2}{\|v_{\mathrm{kin}}\|^2}.
\end{equation*}
\end{proposition}
The following corollary now follows from the observations made in the final paragraph of the preceding section.

\begin{corollary}\label{superluminal}
With the same assumptions as above, the Fermi relative velocity of a radially moving test particle at position $(\tau, \rho)$  within a Fermi coordinate chart satisfies \begin{equation*}
\|v_{\mathrm{Fermi}}\|<\sqrt{-g_{\tau\tau}(\tau, \rho)},
\end{equation*}
and can therefore exceed the central observer's local speed of light ($c=1$) if and only if $-g_{\tau\tau}(\tau, \rho) >1$.
\end{corollary}

Returning to the curvature-normalized coordinates, and exploiting the symmetry of this spacetime through the killing field $\partial/\partial\chi$, the kinematic relative velocity may also be expressed explicitly in terms of $\dot{\chi}_{\mathrm{s}}$.  From $g\left( \tau _{q_{\mathrm{s}} p}u'_{\mathrm{s}},X_p\right) =g\left( u'_{\mathrm{s}},X_{q_{\mathrm{s}}}\right) $ and $g(\tau _{q_{\mathrm{s}} p}u'_{\mathrm{s}},\tau _{q_{\mathrm{s}} p}u'_{\mathrm{s}}) =-1$ we can obtain $\tau _{q_{\mathrm{s}} p}u'_{\mathrm{s}}$, and then find that,
\begin{equation*}
\label{vkingeneral}
v_{\mathrm{kin}} = \frac{\dot{t}_{\mathrm{s}}\sqrt{\frac{a^2(\tau )}{a^2(t_{\mathrm{s}})}-1}+a(\tau )\dot{\chi }_{\mathrm{s}}}{\sqrt{\left( \dot{t}_{\mathrm{s}}\sqrt{\frac{a^2(\tau )}{a^2(t_{\mathrm{s}})}-1}+a(\tau )\dot{\chi }_{\mathrm{s}}\right) ^2+1}}\mathcal{S}_p ,
\end{equation*}
where $\dot{t}_{\mathrm{s}}=\sqrt{a^2(t_{\mathrm{s}})\dot{\chi }_{\mathrm{s}}^2+1}$ by \eqref{dott}.

We conclude this section with examples to illustrate the proposition.
\begin{example}\quad
\begin{enumerate}
\item[a)] For Milne universe, $-g_{\tau\tau}(\tau, \rho)\equiv 1$ and Fermi coordinates are just Minkowski coordinates, so the Fermi chart is global, and all Fermi relative speeds are subluminal.
\item[b)] For the de Sitter universe, $-g_{\tau\tau}(\tau, \rho)=\cos^2(H_0\rho)$, with $H_0\rho <\pi/2$ (see \cite{Klein11,CM,KC3}) where $H_0$ is the Hubble constant. The Fermi chart is valid up to the cosmological horizon of this spacetime. Thus, all Fermi relative velocities are less than the local speed of light.
\item[c)] For the radiation-dominated universe, i.e., for the case that $a(t)=\sqrt{t}$ in \eqref{eqds2}, the Fermi chart is global and
\begin{equation*}
\label{radgtt}
-g_{\tau\tau}(\tau, \rho)=\frac{1}{\sigma}\left(1+\sqrt{\sigma-1}\,\sec^{-1}\sqrt{\sigma}\right)^2,
\end{equation*}
where $\sigma\geq1$ is a parameter that depends on $\rho$ and $\tau$. It may be shown that for any $\tau>0$, the least upper bound of $\rho$ is $\frac{\pi}{2}\tau$ and that $\sqrt{-g_{\tau\tau}(\tau, \rho)}\rightarrow \frac{\pi}{2}$ asymptotically as  $\rho\rightarrow\frac{\pi}{2}\tau$ (see \cite{Klein11}). Thus, Fermi relative speeds can exceed the local speed of light in this spacetime, but are bounded above by $\frac{\pi}{2}$.
\end{enumerate}
\end{example}

\section{Lightlike simultaneity and optical coordinates}
\label{secspecast1}

As in the preceding sections, let $p=(\tau ,0)$ be an event of the central observer $\beta $. An event is \textit{lightlike simultaneous} with $p$ if it lies on the past-pointing horismos $E^-_p$ (which is tangent to the backward light cone at the spacetime point $p$). The vector field,
\begin{equation}
\label{eqY}
Y:=\frac{\partial}{\partial \delta}=\frac{d t}{d \delta}\frac{\partial}{\partial t}+\frac{d \chi}{d \delta}\frac{\partial }{\partial \chi}=-\frac{a(\tau )}{a(t)}\frac{\partial }{\partial t}+\frac{a(\tau )}{a^2(t)}\frac{\partial }{\partial \chi},
\end{equation}
is geodesic, lightlike, and the integral curve $\lambda$ such that $\lambda (0)=p$ is a past-pointing null geodesic, parameterized in such a way that that $\delta$ is the affine distance from $p$ to $\lambda(\delta)$ (see \cite[Proposition 6]{Bolos07}). Let $q_{\ell}=\left( t_{\ell},\chi _{\ell}\right) $ be the unique event of $\beta ' \cap E^-_p$. Then $\lambda $ is the unique geodesic from $p$ to $q_{\ell}$, and so, using \eqref{eqY} we can find a relationship between $\tau$, $t_{\ell}$ and $\chi _{\ell}$:
\begin{equation}
\label{eqt1b}
-\int _{\tau }^{t_{\ell}} \frac{1}{a(t)}\,\mathrm{d}t=\int _0^{\chi _{\ell}} \,\mathrm{d}\chi
\quad \Longrightarrow \quad
\int _{t_{\ell}}^{\tau } \frac{1}{a(t)}\,\mathrm{d}t=\chi _{\ell}.
\end{equation}
From \eqref{eqt1b} we see that $t_{\ell}<\tau$, and more generally, it follows from \eqref{eqY} that $t$ is a decreasing function of affine distance $\delta$. Since we consider only positive coordinate times, $t_{\ell}>0$, and then it is necessary that $0<\chi _{\ell}<\chi _{\ell \mathrm{max}}(\tau )$, where
\begin{equation}
\label{eqchiLmax}
\chi _{\ell \mathrm{max}}(\tau ):=\int _{0}^{\tau } \frac{1}{a(t)}\,\mathrm{d}t
\end{equation}
is the \textit{particle horizon} for the observer $\beta $ at $p$.

In the framework of lightlike simultaneity, it will be convenient to use optical (also called observational) coordinates with respect to the observer $\beta $. Referring to Figure \ref{diagram}, we set the optical coordinates of the point $q_{\ell}=(t_{\ell},\chi _{\ell})$ to be $\left( \tau ,\delta \right)$. From \eqref{eqt1b}, $\tau (t_{\ell},\chi _{\ell})$ is determined implicitly, and differentiation gives,
\begin{equation}
\label{partialtaul}
\frac{\partial \tau }{\partial t_{\ell}}=\frac{a(\tau )}{a(t_{\ell})},\qquad \qquad
\frac{\partial \tau }{\partial \chi _{\ell}}=a(\tau ).
\end{equation}
It follows from \eqref{eqY} that,
\begin{equation}
\label{delta}
\delta=\delta (t_{\ell},\chi _{\ell})=\int _{t_{\ell}}^{\tau (t_{\ell},\chi _{\ell})} \frac{a(t)}{a\left( \tau (t_{\ell},\chi _{\ell})\right) }\,\mathrm{d}t.
\end{equation}
Differentiating \eqref{delta} and using \eqref{partialtaul}, gives
\begin{equation}
\label{partialdeltal}
\frac{\partial \delta }{\partial t_{\ell}}=\frac{a(\tau )}{a(t_{\ell})}-\frac{a(t_{\ell})}{a(\tau )}-\delta \frac{\dot{a}(\tau )}{a(t_{\ell})},\qquad \qquad
\frac{\partial \delta }{\partial \chi _{\ell}}=a(\tau )-\delta \dot{a}(\tau ),
\end{equation}
where $\dot{a}(t)$ is the derivative of $a(t)$. Now using \eqref{partialtaul} and \eqref{partialdeltal}, we may express the Robertson-Walker metric in optical coordinates (with respect to $\beta $) in the form
\begin{equation}
\label{opticalmetric}
\mathrm{d}s^2=\tilde{g}_{\tau\tau}\mathrm{d}\tau ^2+2\mathrm{d}\tau \mathrm{d}\delta\equiv -2\left( 1-\frac{\dot{a}(\tau)}{a(\tau)}\delta -\frac{1}{2}\frac{a^2\left( t_{\ell }\right)}{a^2(\tau )}\right)\mathrm{d}\tau ^2+2\mathrm{d}\tau \mathrm{d}\delta ,
\end{equation}
where $t_{\ell}(\tau ,\delta )$ is given implicitly by \eqref{delta}.

\section{Astrometric and spectroscopic relative velocities}

Based on the previous section, we may write the 4-velocity of the radially moving test particle with worldline $\beta '$ in optical coordinates as, $u'_{\ell}=\dot{\tau }\frac{\partial }{\partial \tau }\vert _{q_{\ell }}+\dot{\delta }\frac{\partial}{\partial \delta }\vert _{q_{\ell }}=(\dot{\tau },\dot{\delta})$, where the overdot represents differentiation with respect to proper time of $\beta '$.

Let $p=(\tau ,0)$ be the event of $\beta $ from which we measure the velocities.  The astrometric relative velocity of $u'_{\ell}$ with respect to $u=(1,0)$ is given by
\begin{equation}
\label{vastopticalA}
v_{\mathrm{ast}}=\frac{d\delta }{d\tau }\mathcal{S}_p=\frac{\dot{\delta }}{\dot{\tau }}\mathcal{S}_p.
\end{equation}
From \eqref{opticalmetric}, \eqref{vastopticalA} and the requirement that $g\left( u'_{\ell},u'_{\ell}\right) =-1$, we obtain,
\begin{equation}
\label{vastoptical}
v_{\mathrm{ast}}=\frac{\dot{\delta }}{\dot{\tau }}\mathcal{S}_p = \frac{1}{2}\left(-\tilde{g}_{\tau\tau}-\frac{1}{\dot{\tau}^2}\right)\mathcal{S}_p.
\end{equation}
There is no upper bound for $\| v_{\mathrm{ast}}\|$ in the case of a radially approaching test particle (i.e., for the case $d\delta/d\tau<0$) because $\dot{\tau}$ can be chosen to be arbitrarily close to zero in \eqref{vastoptical}. However, a sharp upper bound for the case of a radially receding test particle is given by $\| v_{\mathrm{ast}}\| < -\tilde{g}_{\tau\tau}/2$, where the right side of this inequality is the relative speed, $d\delta/d\tau$, of a distant radially receding photon. Since it follows from \eqref{opticalmetric} that $-\tilde{g}_{\tau\tau}<2$ when $\dot{a}(\tau)\geq0$, we have the following general result.

\begin{proposition}
\label{subluminal}
In any expanding Robertson-Walker spacetime, the astrometric relative velocity of a radially receding test particle is always less than the central observer's local speed of light ($c=1$).
\end{proposition}

The spectroscopic relative velocity, discussed briefly in the introduction, is defined in a way analogous to the definition of the kinematic relative velocity given in \eqref{vk1}.  However, for the case of radial motion, the following equivalent formula, more convenient for our purposes, was deduced in \cite{Bolos07}:
\begin{equation}
\label{vspecformula}
v_{\mathrm{spec}}=\frac{\left(\dfrac{\nu'}{\nu }\right)^2-1}{\left(\dfrac{\nu'}{\nu}\right)^2+1}\,\mathcal{S}_p,
\end{equation}
where $\nu $, $\nu '$ are the frequencies observed by $u$, $u'_{\ell }$, respectively, of a photon emitted from the spacetime point $q_{\ell }$. It is clear from \eqref{vspecformula} that $\|v_{\mathrm{spec}}\|$ must be less than 1.  The frequency ratio is given by,
\begin{equation}
\label{doppleroptic}
\frac{\nu '}{\nu }=\frac{g(\textrm{P}_{q_{\ell }},u'_{\ell })}{g(\textrm{P}_p,u)},
\end{equation}
where $\textrm{P}:=-\partial /\partial \delta $ is the 4-momentum tangent vector field of the emitted photon.
Using \eqref{opticalmetric} in \eqref{doppleroptic}, we find,
\begin{equation}
\label{doppleroptic2}
\frac{\nu '}{\nu }=\dot{\tau},
\end{equation}
and thus, applying \eqref{doppleroptic2} in \eqref{vspecformula}, we obtain,
\begin{equation}
\label{vspecinoptical}
v_{\mathrm{spec}}=\frac{\dot{\tau}^2-1}{\dot{\tau}^2+1}\mathcal{S}_p.
\end{equation}

In order to find expressions for the astrometric and spectroscopic relative velocities in terms of curvature-normalized coordinates, we make use of \eqref{partialtaul} to obtain
\begin{equation}
\label{taudot}
\dot{\tau}=\frac{\partial \tau }{\partial t_{\ell}}\dot{t}_{\ell}+\frac{\partial \tau }{\partial \chi _{\ell}}\dot{\chi }_{\ell} = a(\tau )\left( \sqrt{\dot{\chi }_{\ell}^2+a^{-2}(t_{\ell})}+\dot{\chi }_{\ell}\right) ,
\end{equation}
where we have used the identification, $\dot{t}_{\ell}=\sqrt{a^2(t_{\ell})\dot{\chi }_{\ell}^2+1}$, which follows from \eqref{dott}. Combining this last expression with \eqref{vspecinoptical} yields,
\begin{equation*}
\label{vspecgeneral2}
v_{\mathrm{spec}} =\frac{a^2(\tau )\left( \sqrt{\dot{\chi }_{\ell}^2+a^{-2}(t_{\ell})}+\dot{\chi }_{\ell}\right) ^2-1}{a^2(\tau )\left( \sqrt{\dot{\chi }_{\ell}^2+a^{-2}(t_{\ell})}+\dot{\chi }_{\ell}\right) ^2+1}\mathcal{S}_p.
\end{equation*}
Similarly, combining \eqref{taudot} with \eqref{vastoptical} gives,
\begin{equation*}
\label{vastgeneral3}
v_{\mathrm{ast}}=\left(1- \frac{\dot{a}(\tau )}{a(\tau )}\delta-\frac{a^2(t_{\ell})}{2a^2(\tau )}
\left[1+\left( \sqrt{a^2(t_{\ell})\dot{\chi }_{\ell}^2+1}+a(t_{\ell})\dot{\chi }_{\ell}\right) ^{-2}\right]\right) \mathcal{S}_p.
\end{equation*}
Moreover, combining \eqref{vspecinoptical} with \eqref{vastoptical} yields a relationship between the spectral and astrometric relative velocities:
\begin{equation}
\label{astro/spec}
v_{\mathrm{spec}}=\frac{1+\tilde{g}_{\tau\tau}(\tau,\delta) \pm \| v_{\mathrm{ast}}\|}{1-\tilde{g}_{\tau\tau}(\tau,\delta) \mp \| v_{\mathrm{ast}}\|}\mathcal{S}_p,
\end{equation}
where in the case that $d\delta/d\tau>0$ (i.e., in the case of a receding test particle), the positive sign in the numerator and negative sign in the denominator are chosen, and the opposite choices of signs are taken when $d\delta/d\tau<0$ (i.e., in the case of an approaching test particle).

Now using \eqref{astro/spec} we may formulate the following proposition.

\begin{proposition}\label{findmetric}
For a Robertson-Walker spacetime with scale factor $a(t)$ that is a smooth, increasing function of $t$, a measurement of the astrometric and spectroscopic speeds of a receding test particle  relative to the comoving observer  determine the metric tensor element $\tilde{g}_{\tau\tau}$ in optical coordinates at the spacetime point of the particle via,
\begin{equation*}
\tilde{g}_{\tau\tau}(\tau,\delta)=\frac{1- \| v_{\mathrm{ast}}\|}{1+\|v_{\mathrm{spec}}\|}\|v_{\mathrm{spec}}\|-\frac{1+ \| v_{\mathrm{ast}}\|}{1+\|v_{\mathrm{spec}}\|}.
\end{equation*}
\end{proposition}

\section{Fermi coordinate charts for power law cosmologies}
\label{fermi}

It was shown in \cite{Klein11} that the Fermi coordinate chart for a comoving observer covers the entire Robertson-Walker spacetime, provided the scale factor $a(t)$ is smooth, increasing, unbounded, and for all $t>0$, $\ddot{a}(t)\leq0$, i.e., for non inflationary cosmologies (with or without a big bang).

Cosmologies with scale factors of the form $a(t)=t^{\alpha}$, with $0<\alpha\leq1$ fall within this category and have global Fermi coordinates for comoving observers.  Included are the radiation dominated universe ($\alpha=1/2$) and the matter dominated universe ($\alpha=2/3$).

Here we consider scale factors of the form $a(t)=t^{\alpha}$, with $\alpha>1$.  Power law cosmologies with $\alpha>1$ have been used to model dark energy, and astronomical measurements have been made to support their consideration (see \cite{power}). These cosmologies are inflationary ($\ddot{a}>0$), and have the additional property that they contain spacetime points from which the central observer can never receive light signals, i.e., these cosmologies include \textit{cosmological event horizons}.  The $\chi$-coordinate at time $t_{\mathrm{s}}$ of the event horizon is given by,
\begin{equation*}
\chi_{\mathrm{horiz}}(t_{\mathrm{s}}):= \int_{t_{\mathrm{s}}}^{\infty}\frac{\mathrm{d}t}{a(t)},
\end{equation*}
or more explicitly,
\begin{equation*}
\chi_{\mathrm{horiz}}(t_{\mathrm{s}}) = \int_{t_{\mathrm{s}}}^{\infty}\frac{\mathrm{d}t}{t^\alpha} = \frac{t_{\mathrm{s}}^{1-\alpha}}{\alpha-1} < +\infty.
\end{equation*}
We note that in contrast to the case $0<\alpha<1$, there is no particle horizon when $\alpha>1$, i.e., the integral in \eqref{eqchiLmax} is infinite, so that the astrometric and spectroscopic relative velocities are well-defined for arbitrarily large $\chi$-coordinates. Let,
\begin{equation*}
\label{V}
\mathcal{V} := \left\{ (t, \chi) : t>0 \text{ and } 0<\chi < \chi_{\mathrm{horiz}}(t)\right\}.
\end{equation*}
Intuitively, $\mathcal{V}$ is the set of all eventually observable events. For $0<\alpha<1$,  $\mathcal{V}=\left] 0,+\infty\right[ \times \left] 0,+\infty\right[ $ represents all spacetime points distant from the central observer.

We will show that the set of spacetime points with coordinates in $\mathcal{V}$ is a maximal chart for Fermi coordinates for a central observer in an inflationary power law cosmology. For that purpose, we extract the integral from \eqref{eqt1} and define,
\begin{equation}
\label{key}
\chi_{\mathrm{s}}(\tau):=\int _{t_{\mathrm{s}}}^{\tau } \frac{a(\tau )}{a(t)}\frac{1}{\sqrt{a^2(\tau )-a^2(t)}}\,\mathrm{d}t.
\end{equation}
In geometric terms, $\chi_{\mathrm{s}}(\tau)$ is the value of the $\chi$-coordinate of the point with $t$-coordinate $t_{\mathrm{s}}$ on the spacelike geodesic orthogonal to $\beta$ with initial point $(\tau,0)$ on the central observer's worldline.  With the change of variables, $\tilde{\sigma}=\left(a(\tau)/a(t)\right)^{2}$ (with $\tau$ held fixed), expression \eqref{key} becomes,
\begin{equation}
\label{key2}
\chi_{\mathrm{s}}(\tau)=\frac{1}{2}\int_{1}^{\sigma(\tau)}\dot{b}\left(\frac{a(\tau)}{\sqrt{\tilde{\sigma}}}\right)\frac{1}{\sqrt{\tilde{\sigma}}\sqrt{\tilde{\sigma}-1}}\,\mathrm{d}\tilde{\sigma},
\end{equation}
where $b(t)$ is the inverse function of $a(t)$, (so that $b(a(t))=t$), and,
\begin{equation}
\label{defsig}
\sigma(\tau):=\left(\dfrac{a(\tau)}{a(t_{\mathrm{s}})}\right)^2.
\end{equation}

It follows from \eqref{eqX} that $t_{\mathrm{s}}\leq\tau$ and that $t_{\mathrm{s}}$ decreases with increasing proper distance along spacelike geodesics orthogonal to the central observer worldline $\beta$. Then  $\sigma(\tau)\geq1$ and for a given initial point $(\tau,0)$, $\sigma$ may be used as a (non affine) parameter for this geodesic.

It was shown in \cite{Klein11} that for a class of scale factors that includes the inflationary power laws considered here, the spacelike geodesic $\psi$ depicted in Figure \ref{diagram}, may be parameterized as, $\psi(\sigma)=(t(\tau, \sigma), \chi(\tau, \sigma))$, where $\sigma \geq 1$, and,
\begin{equation}
\label{toftau}
t(\tau, \sigma)=b\left(\dfrac{a(\tau)}{\sqrt{\sigma}}\right) ,
\end{equation}
\begin{equation}
\label{chi}
\chi(\tau,\sigma)=\frac{1}{2}\int_{1}^{\sigma}\dot{b}\left(\frac{a(\tau)}{\sqrt{\tilde{\sigma}}}\right)\frac{1}{\sqrt{\tilde{\sigma}}\sqrt{\tilde{\sigma}-1}}\,\mathrm{d}\tilde{\sigma}.
\end{equation}
Our strategy to prove that the set of spacetime points with coordinates in $\mathcal{V}$ is a chart for Fermi coordinates involves first showing that $(\tau,\sigma)$ are coordinates on that chart.  To carry this through, we require a sequence of lemmas.

\begin{lemma}\label{bound}
For $a(t)=t^{\alpha}$ with $\alpha > 1$, and any $\tau>t_{\mathrm{s}}>0$,
\begin{equation*}
\chi_{\mathrm{s}}(\tau) < \chi_{\mathrm{horiz}}(t_{\mathrm{s}}).
\end{equation*}
\end{lemma}

\begin{proof}
Substituting $a(t)=t^{\alpha}$ into \eqref{key2} gives,
\begin{equation}
\label{fraction}
\chi_{\mathrm{s}}(\tau)=\dfrac{1}{2\alpha\tau^{\alpha-1}} \int_{1}^{\sigma(\tau)}\frac{1}{{\tilde{\sigma}^{1/2\alpha}}\sqrt{\tilde{\sigma}-1}}\,\mathrm{d}\tilde{\sigma}.
\end{equation}
The substitution $x^{2\alpha}=\tilde{\sigma}$ then gives,
\begin{equation}
\label{fraction2}
\chi_{\mathrm{s}}(\tau)=\dfrac{1}{\tau^{\alpha-1}} \int_{1}^{\tau/t_{\mathrm{s}}}\frac{1}{x}\frac{x^{2\alpha-1}}{\sqrt{x^{2\alpha}-1}}\,\mathrm{d}x.
\end{equation}
Integration by parts in \eqref{fraction2} yields,
\begin{equation*}
\label{fraction3}
\begin{split}
\chi_{\mathrm{s}}(\tau)=&\dfrac{1}{\alpha\tau^{\alpha-1}}\left[\frac{\sqrt{(\tau/t_{\mathrm{s}})^{2\alpha}-1}}{\tau/t_{\mathrm{s}}} +\int_{1}^{\tau/t_{\mathrm{s}}}\frac{\sqrt{x^{2\alpha}-1}}{x^{2}}\,\mathrm{d}x\right]\\
<&\,\dfrac{1}{\alpha\tau^{\alpha-1}}\left[\left(\frac{\tau}{t_{\mathrm{s}}}\right)^{\alpha-1}+\int_{1}^{\tau/t_{\mathrm{s}}}x^{\alpha-2}\,\mathrm{d}x\right]\\
<&\,\dfrac{1}{\alpha\tau^{\alpha-1}}\left[\left(\frac{\tau}{t_{\mathrm{s}}}\right)^{\alpha-1}+\frac{1}{\alpha-1}\left(\frac{\tau}{t_{\mathrm{s}}}\right)^{\alpha-1}\right]\\
=&\frac{t_{\mathrm{s}}^{1-\alpha}}{\alpha-1}=\chi_{\mathrm{horiz}}(t_{\mathrm{s}}).
\end{split}
\end{equation*}
\end{proof}

\begin{lemma}\label{horizon}
For $a(t)=t^{\alpha}$ with $\alpha > 1$, and any $t_{\mathrm{s}}>0$,
\begin{equation*}
\lim_{\tau\to+\infty}\chi_{\mathrm{s}}(\tau) =\chi_{\mathrm{horiz}}(t_{\mathrm{s}}).
\end{equation*}
\end{lemma}

\begin{proof}
From L'Hôpital's and Leibniz's rules applied to \eqref{fraction}, we have,
\begin{equation*}
\lim_{\tau\to+\infty}{\chi_{\mathrm{s}}(\tau)}=\lim_{\tau\to+\infty}{\left(\frac{t_{\mathrm{s}}^{1-\alpha}}{\alpha-1}\right)\left(\frac{\tau^\alpha}{\sqrt{\tau^{2\alpha}-t_{\mathrm{s}}^{2\alpha}}}\right)
}= \frac{t_{\mathrm{s}}^{1-\alpha}}{\alpha-1} = \chi_{\mathrm{horiz}}(t_{\mathrm{s}}).
\end{equation*}
\end{proof}

\begin{lemma}\label{increase}
For $a(t)=t^{\alpha}$ with $\alpha > 1$, and any $\tau>t_{\mathrm{s}}>0$,
\begin{equation*}
\dfrac{d\chi_{\mathrm{s}}}{d\tau}(\tau)>0.
\end{equation*}
\end{lemma}

\begin{proof}
For convenience, rewrite \eqref{fraction} as $\chi_{\mathrm{s}}(\tau)=\dfrac{f(\tau)}{g(\tau)}$ where,
\begin{equation*}
f(\tau):= \int_{1}^{\sigma(\tau)}\frac{1}{{\tilde{\sigma}^{1/2\alpha}}\sqrt{\tilde{\sigma}-1}}\,\mathrm{d}\tilde{\sigma},\quad\text{and}\quad g(\tau) := 2\alpha\tau^{\alpha-1}.
\end{equation*}
By the quotient rule, $\dfrac{d\chi_{\mathrm{s}}}{d\tau}>0$ if and only if
\begin{equation}
\label{eq:fg}
\chi_{\mathrm{s}}(\tau)=\frac{f(\tau)}{g(\tau)}<\frac{f'(\tau)}{g'(\tau)},
\end{equation}
By Leibniz's rule, the quotient on the right in \eqref{eq:fg} is,
\begin{equation}
\label{eq:fg2}
\dfrac{f'(\tau)}{g'(\tau)} = \left(\frac{t_{\mathrm{s}}^{1-\alpha}}{\alpha-1}\right)\left( \dfrac{\tau^{\alpha}}{\sqrt{\tau^{2\alpha}-t_{\mathrm{s}}^{2\alpha}}} \right).
\end{equation}
Since the last term on the right in \eqref{eq:fg2} is always greater than 1, we have,
\begin{equation*}
\dfrac{f'(\tau)}{g'(\tau)}> \frac{t_{\mathrm{s}}^{1-\alpha}}{\alpha-1} = \chi_{\mathrm{horiz}}(t_{\mathrm{s}}),
\end{equation*}
Therefore $\dfrac{d\chi_{\mathrm{s}}}{d\tau}>0$ for any $\tau$ such that $\chi_{\mathrm{s}}(\tau)<\chi_{\mathrm{horiz}}(t_{\mathrm{s}})$.  The result now follows from Lemma \ref{bound}.
\end{proof}

\begin{lemma}
\label{powerlemma}
For $a(t)=t^{\alpha}$, the map $F : \left] 0, \infty \right[ \times \left] 1, \infty \right[ \rightarrow \mathcal{V}$ given by
\begin{equation*}
\label{Function}
F(\tau,\sigma):=\left(t(\tau,\sigma), \chi(\tau,\sigma)\right),
\end{equation*}
where the functions $t$ and $\chi$ are defined by \eqref{toftau} and \eqref{chi}, respectively, is a diffeomorphism.
\end{lemma}

\begin{proof}
Let $(t_{\mathrm{s}}, \chi_{\mathrm{s}}) \in \mathcal{V}$ be arbitrary but fixed.  To prove that $F$ is a bijection, we must show that there exists a unique pair $(\tau_0, \sigma_0) \in \left] 0, +\infty \right[ \times \left] 1, +\infty \right[ $ such that $F(\tau_0,\sigma_0)=(t_{\mathrm{s}}, \chi_{\mathrm{s}})$.  From \eqref{toftau}, it follows that $\sigma_0$ is uniquely determined by $\tau_0$ and,

\begin{equation*}
\sigma_0= \left( \dfrac{a(\tau_0)}{a(t_{\mathrm{s}})} \right)^2.
\end{equation*}
So it remains only to find $\tau_0$.  To that end, let $\sigma(\tau)$ be given by \eqref{defsig}.  It then follows from \eqref{key2} and \eqref{chi} that
\begin{equation*}
\chi(\tau,\sigma(\tau))=\chi_{\mathrm{s}}(\tau).
\end{equation*}
Since by assumption, $\chi_{\mathrm{s}}<\chi_{\mathrm{horiz}}(t_{\mathrm{s}})$, it follows from Lemmas \ref{horizon} and \ref{increase} that there is a unique $\tau_0>t_{\mathrm{s}}$ such that $\chi(\tau_{0},\sigma(\tau_{0}))=\chi((\tau_{0},\sigma_{0})=\chi_{\mathrm{s}}$.  Thus, $F$ is a bijection.

The Jacobian determinant $J(\tau,\sigma)$ for the transformation $F$ was calculated in \cite{Klein11} for a general class of scale factors including the power law scale factors considered here, and is given by,
\begin{equation}
\label{jacobian}
J(\tau,\sigma)=\frac{\dot{a}(\tau)}{2 \sigma}\dot{b} \left(\frac{a(\tau)}{\sqrt{\sigma}}\right) \left(\frac{\dot{b} \left(\frac{a(\tau)}{\sqrt{\sigma}}\right)}{\sqrt{\sigma-1}} +\frac{a(\tau)}{2\sqrt{\sigma}}\int_{1}^{\sigma} \frac{\ddot{b}\left(\frac{a(\tau)}{\sqrt{\tilde{\sigma}}}\right)}{\tilde{\sigma}\sqrt{\tilde{\sigma}-1}}\,\mathrm{d}\tilde{\sigma}\right).
\end{equation}
Thus, from \eqref{jacobian},
\begin{equation}
\label{joftau}
J\left(\tau,\sigma(\tau)\right)=\frac{a(\tau)\dot{a}(\tau)}{2 \sigma(\tau)^{3/2}}\dot{b}\left(\frac{a(\tau)}{\sqrt{\sigma(\tau)}}\right)
 \left[\frac{\dot{b}\left(\frac{a(\tau)}{\sqrt{\sigma(\tau)}}\right)\sqrt{\sigma(\tau)}}{a(\tau)\sqrt{\sigma(\tau)-1}} +\frac{1}{2}\int_{1}^{\sigma(\tau)}\frac{\ddot{b}\left(\frac{a(\tau)}{\sqrt{\tilde{\sigma}}}\right)}{\tilde{\sigma}\sqrt{\tilde{\sigma}-1}}\,\mathrm{d}\tilde{\sigma}\right].
\end{equation}
The first term in the square brackets in \eqref{joftau} may be rewritten:
\begin{equation}
\label{bdot}
\frac{\dot{b}\left(\frac{a(\tau)}{\sqrt{\sigma(\tau)}}\right)\sqrt{\sigma(\tau)}}{a(\tau)\sqrt{\sigma(\tau)-1}}  = \dfrac{a(\tau)\dot{b}(a(t_{\mathrm{s}}))}{a(t_{\mathrm{s}})^2\sqrt{\sigma(\tau)}\sqrt{\sigma(\tau)-1}}.
\end{equation}
Now, applying Leibniz's rule to \eqref{key2} and \eqref{defsig} yields,
\begin{equation}
\label{eq:3}
\dfrac{d\chi(\tau)}{d\tau}=\dot{a}(\tau)\left[\dfrac{a(\tau)\dot{b}(a(t_{\mathrm{s}}))}{a(t_{\mathrm{s}})^2\sqrt{\sigma(\tau)}\sqrt{\sigma(\tau)-1}} + \dfrac{1}{2}\int_{1}^{\sigma}\frac{\ddot{b}\left(\frac{a(\tau)}{\sqrt{\tilde{\sigma}}}\right)}{\tilde{\sigma}\sqrt{\tilde{\sigma}-1}}\,\mathrm{d}\tilde{\sigma}\right].
\end{equation}
Combining \eqref{joftau}, \eqref{bdot} and \eqref{eq:3} gives,
\begin{equation}
\label{eq:4}
J(\tau, \sigma(\tau))=\dfrac{a(\tau)\dot{b}(a(t_{\mathrm{s}}))}{2\sigma(\tau)^{3/2}}\dfrac{d\chi(\tau)}{d\tau}.
\end{equation}
Thus, applying Lemma \ref{increase} in \eqref{eq:4}, $J(\tau, \sigma(\tau))>0$ for all $\tau$.  Now given any $\tau>0$ and $\sigma>1$ there exists a positive $t_{\mathrm{s}}< \tau$ such that $\sigma(\tau)=(\tau/t_{\mathrm{s}})^{2\alpha}=\sigma$. Therefore $J(\tau, \sigma)>0$ for all $(\tau,\sigma)$, and by the inverse function theorem, $F$ is a diffeomorphism.
\end{proof}

To finish the construction of Fermi coordinates, we return to \eqref{eqrho} in the form,
\begin{equation}
\label{eqrho2}
\rho=\int _{t_{\mathrm{s}}}^{\tau}\frac{a(t)}{\sqrt{a^2(\tau)-a^2(t)}}\,\mathrm{d}t,
\end{equation}
which gives the proper distance along the spacelike geodesic ($\psi$ in Figure \ref{diagram}) orthogonal to $\beta$, from the initial point $(\tau,0)\in\beta$ to the unique point whose $t$-coordinate is $t_{\mathrm{s}}$.  The radius $\rho_{\mathcal{M}_{\tau}}$ of the Fermi slice, $\mathcal{M}_{\tau}$, at proper time $\tau$ of the central observer is obtained by replacing $t_{\mathrm{s}}$ by zero in \eqref{eqrho2}.  The result for $a(t)=t^{\alpha}$ is,
\begin{equation}
\label{Mradius}
\rho_{\mathcal{M}_{\tau}}=\rho_{\mathcal{M}_{\tau}}(\alpha)=\frac{\sqrt{\pi }\,\Gamma \left( \frac{1+\alpha }{2\alpha }\right) }{\Gamma \left( \frac{1}{2\alpha }\right)}\tau ,
\end{equation}
which holds for all $\alpha>0$ (see \cite{Klein11}).
The change of variables, $\tilde{\sigma}=\left(a(\tau)/a(t)\right)^{2}$, applied to the integral in \eqref{eqrho2} results in the following formula for the Fermi coordinate $\rho$, which also appears in \cite{Klein11}:
\begin{equation}
\label{rhotausigma}
\rho = \rho_{\tau}(\sigma) = \dfrac{a(\tau)}{2}\int_{1}^{\sigma}\dot{b}\left(\frac{a(\tau)}{\sqrt{\tilde{\sigma}}}\right)\frac{1}{\tilde{\sigma}^{3/2}\sqrt{\tilde{\sigma}-1}}\,\mathrm{d}\tilde{\sigma}.
\end{equation}
It is clear from \eqref{rhotausigma} that for a fixed value of $\tau$, $\rho_{\tau}(\sigma)$ is an increasing function of $\sigma$, and therefore has an inverse function, $\sigma_{\tau}(\rho)$. Define,
\begin{equation*}
\mathcal{U}_{\mathrm{Fermi}} := \left\{ (\tau, \rho) : \tau >0 \text{ and } 0< \rho < \rho_{\mathcal{M}_{\tau}}\right\},
\end{equation*}
and let $G(\tau, \sigma) := (\tau, \rho(\sigma))$.  Then $G$ is a diffeomorphism with inverse $G^{-1}(\tau, \rho) = (\tau, \sigma_{\tau}(\rho))$.  Define $H: \mathcal{V} \to \mathcal{U}_{\mathrm{Fermi}}$ by
\begin{equation*}
H(t, \chi) := G \circ F^{-1}(t,\chi).
\end{equation*}
Then $H$ is a diffeomorphism.  Taking into consideration that $\chi$ and $\rho$ are both positive, radial coordinates, we can now state the main result of this section:

\begin{theorem}
For a comoving observer in a Robertson-Walker cosmology whose scale factor is $a(t)=t^{\alpha}$ with $\alpha > 1$, the maximal domain of Fermi coordinates  is the set of all spacetime points whose curvature coordinates take values in $\mathcal{V}$, i.e., Fermi coordinates extend to the cosmological event horizon.  The range of Fermi coordinates is $\mathcal{U}_{\mathrm{Fermi}}$.
\end{theorem}

\section{Comparisons of relative velocities in cosmologies with power law scale factors}
\label{sec:5}

Consider a test particle comoving with the Hubble flow, $\beta '(\tau')=(\tau',\chi)$, where $\chi >0$ is constant. Referring to Figure \ref{diagram}, we have that $q_{\mathrm{s}}=( t_{\mathrm{s}},\chi) $ and $q_{\ell}=\left( t_{\ell},\chi \right) $; moreover $u'_{\mathrm{s}}=\left. \frac{\partial }{\partial t}\right| _{q_{\mathrm{s}}}=(1,0)$ and $u'_{\ell}=\left. \frac{\partial }{\partial t}\right| _{q_{\ell}}=(1,0)$.

In \eqref{eqt1}, $t_{\mathrm{s}}$ is implicitly defined as a function of $(\tau ,\chi)$, and similarly in \eqref{eqt1b}, $t_{\ell}$ is implicitly defined as a function of $( \tau ,\chi) $. From now on, it will be convenient to regard not only  $t_{\mathrm{s}}$ and $t_{\ell}$ as functions of $(\tau ,\chi)$, but also the four relative velocities.  However, it is important to recognize that in this context $\chi$ is a parameter that labels a comoving test particle (with fixed coordinate $\chi$), and $\tau$ is the time of observation by the central observer $\beta $.  The relative velocities are vectors in the tangent space of the point $p=(\tau,0)$ for test particles with coordinates $(t_{\mathrm{s}}, \chi)$ in the case of the Fermi and kinematic relative velocities, and with coordinates $(t_{\ell}, \chi)$ in the case of the astrometric and spectroscopic relative velocities. Since all the velocities are proportional to $\mathcal{S}_p$ and in the same direction, we will find expressions only for the moduli of the relative velocities.

In \cite{BolosKlein}, comoving test particles in cosmologies with a variety of scale factors were studied. These include power law scale factors of the form
\[
a(t)=t^{\alpha}
\]
with $0<\alpha \leq 1$. Here, we relax that restriction and allow $\alpha >1$.

In preparation for the study of the relative velocities, it is convenient to define a parameter
\begin{equation}
\label{hubblespeed}
v=v(\tau ,\chi ):=\dot{a}(\tau )\chi =\alpha \chi \tau ^{\alpha -1}.
\end{equation}
This parameter will be useful in the description of the relative velocities because their moduli depend on $(\tau,\chi)$ by means of $v$. In \eqref{hubblespeed}, the overdot represents differentiation with respect to $\tau $, and $v$ is the \textit{Hubble speed} of a comoving test particle with curvature-normalized coordinates $(\tau ,\chi )$.

\subsection{Spacelike simultaneity}

For notational convenience in this subsection, let,
\begin{equation}
\label{calpha1}
C_{\alpha }:=\frac{\sqrt{\pi }\,\Gamma \left( \frac{1+\alpha }{2\alpha }\right) }{\Gamma \left( \frac{1}{2\alpha }\right)}.
\end{equation}
It follows from \eqref{Mradius} and \eqref{calpha1} that for any $\alpha >0$,
\begin{equation}
\label{calpha2}
C_{\alpha }=\frac{\rho_{\mathcal{M}_{\tau}}(\alpha)}{\tau},
\end{equation}
where $\rho_{\mathcal{M}_{\tau}}(\alpha)$ is the proper radius of the spaceslice $\mathcal{M}_{\tau}$ for scale factor $a(t)=t^{\alpha}$.

From \eqref{eqchiSmax}, it follows that $\chi _{\mathrm{smax}}(\tau )=+\infty $ for $\alpha>1$, and so for this case, $\chi $ and $v$ have no upper bounds. In contrast, $\chi _{\mathrm{smax}}(\tau )=\frac{\tau ^{1-\alpha}}{1-\alpha}C_{\alpha}$ is finite and $v$ is bounded by $v_{\mathrm{smax}}:=\frac{\alpha}{1-\alpha}C_{\alpha }$ for $0<\alpha<1$ (see \cite{BolosKlein}).

By \eqref{eqt1} we have,
\begin{equation}
\label{eq:tspsf}
\left( \frac{t_{\mathrm{s}}}{\tau }\right) ^{1-\alpha } \,\!_2F_1\left( \frac{1}{2},\frac{1-\alpha }{2\alpha };\frac{1+\alpha }{2\alpha };\left( \frac{t_{\mathrm{s}}}{\tau }\right) ^{2\alpha }\right) = C_{\alpha }+\frac{\alpha -1}{\alpha}v,
\end{equation}
where $ _2F_1(\cdot ,\cdot ;\cdot ;\cdot )$ is the Gauss hypergeometric function.

\begin{figure}[tbp]
\begin{center}
\includegraphics[width=1\textwidth]{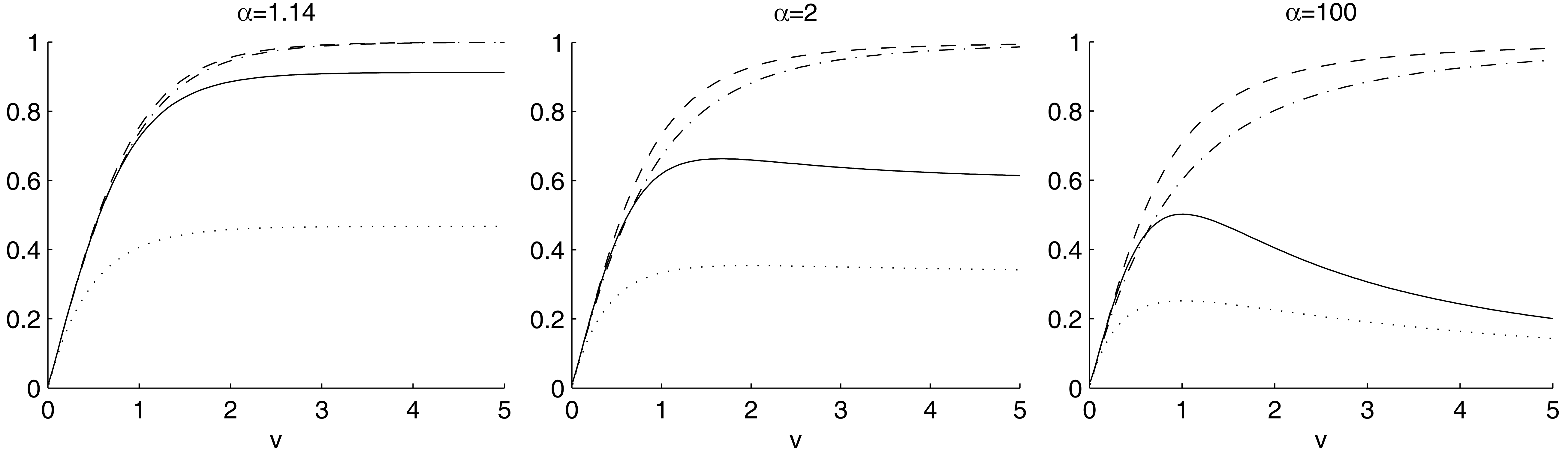}
\end{center}
\caption{Instant comparison of the moduli of kinematic (dashed), Fermi (solid), spectroscopic (dot-dashed) and astrometric (dotted) relative velocities with respect to the parameter $v=\alpha \chi \tau^{\alpha-1}$, in a universe with power scale factor $a(t)=t^{\alpha}$ for $\alpha>1$.}
\label{alphagt1}
\end{figure}

Define the function $F_{\alpha }(z) := z^{\frac{1-\alpha }{\alpha }} \, _2F_1\left( \frac{1}{2},\frac{1-\alpha }{2\alpha };\frac{1+\alpha }{2\alpha };z^2\right) $
where $0<z<1$ and $\alpha >0$, $\alpha\neq 1$\footnote{The case $\alpha=1$ corresponds with the Milne universe and it is studied in \cite{BolosKlein}.}. It is bijective and by \eqref{eq:tspsf},
\begin{equation}
\label{eqt1Galpha}
t_{\mathrm{s}}(\tau ,\chi ) =G_{\alpha } (v)\tau,
\end{equation}
with $G_{\alpha }(v):=\left( F_{\alpha }^{-1}\left( C_{\alpha }+\frac{\alpha -1}{\alpha}v\right) \right)^{1/\alpha }$,
where the superscript $^{-1}$ denotes the inverse function. Using techniques analogous to those used in \cite{BolosKlein}, from \eqref{eqt1Galpha} we find:
\begin{proposition}\label{powerVFermi}
The kinematic and Fermi speeds of a comoving test particle relative to a comoving central observer in a Robertson-Walker cosmology with scale factor $a(t)=t^{\alpha}$ and $\alpha>0$, $\alpha \neq 1$ are given by
\begin{equation}
\label{powerVkinetic}
\| v_{\mathrm{kin}}\| = \sqrt{1-G_{\alpha }^{2\alpha }(v)},
\end{equation}
and
\begin{equation}
\label{eq:vFermi3psf}
\| v_{\mathrm{Fermi}}\| =G_{\alpha }^{1-\alpha }(v)\sqrt{1-G_{\alpha }^{2\alpha }(v)}-\frac{\alpha -1}{\alpha}\left( 1-G_{\alpha }^{2\alpha }(v)\right) v,
\end{equation}
where $v$ is the parameter given by \eqref{hubblespeed}. Moreover, from \eqref{powerVkinetic}, \eqref{eq:vFermi3psf}, and taking into account \eqref{calpha2}, we have,
\begin{equation*}
\lim_{v\rightarrow v_{\mathrm{smax}}}\|v_{\mathrm{kin}}\|=1
\end{equation*}
and
\begin{equation}
\label{velradius}
\lim_{v\rightarrow v_{\mathrm{smax}}}\|v_{\mathrm{Fermi}}\|=C_{\alpha}=\frac{\rho_{\mathcal{M}_{\tau}}}{\tau},
\end{equation}
where $v_{\mathrm{smax}}=\frac{\alpha}{1-\alpha}C_{\alpha}$ if $0<\alpha<1$, and $v_{\mathrm{smax}}=+\infty$ if $\alpha>1$.
\end{proposition}
It follows from \eqref{velradius} that the limiting Fermi speeds exceed $1$ only for $0<\alpha<1$.  More generally, it follows from \eqref{eq:vFermi3psf} that $\|v_{\mathrm{Fermi}}\|<1$ for any comoving particle when $\alpha>1$, i.e., there are no superluminal Fermi velocities for inflationary power law scale factors.
Examples are presented in Figure \ref{alphagt1}.

\subsection{Lightlike simultaneity}

From \eqref{eqchiLmax}, it follows that $\chi _{\ell \mathrm{max}}(\tau )=+\infty $ for $\alpha>1$, and so for this case, $\chi$ and $v$ have no upper bounds in the framework of lightlike simultaneity. In contrast, $\chi _{\ell \mathrm{max}}(\tau )=\frac{\tau ^{1-\alpha }}{1-\alpha }$  is finite and $v$ is bounded by $v_{\ell \mathrm{max}}:=\frac{\alpha }{1-\alpha }$ for $0<\alpha<1$ (see \cite{BolosKlein}).

By \eqref{eqt1b}
\begin{equation}
\label{eq:tlpsf}
t_{\ell }(\tau ,\chi )=\left( 1-\frac{1-\alpha }{\alpha }v\right) ^{\frac{1}{1-\alpha }}\tau .
\end{equation}
Using techniques analogous to those used in \cite{BolosKlein}, from \eqref{eq:tlpsf} we find:
\begin{proposition}\label{powerlight}
The spectroscopic and astrometric speeds of a comoving test particle relative to a comoving central observer in a Robertson-Walker cosmology with scale factor $a(t)=t^{\alpha}$ and $\alpha>0$, $\alpha \neq 1$ are given by
\begin{equation}
\label{eq:vspecpsf}
\| v_{\mathrm{spec}}\| =\frac{1-\left( 1+\frac{\alpha -1}{\alpha }v\right) ^{\frac{2\alpha }{1-\alpha }}}{1+\left( 1+\frac{\alpha -1}{\alpha }v\right) ^{\frac{2\alpha }{1-\alpha }}},
\end{equation}
and
\begin{equation}
\label{eq:vastpsf}
\| v_{\mathrm{ast}}\|  =
\frac{1}{1+\alpha }\left( 1-\left( 1+\frac{\alpha -1}{\alpha }v \right) ^{\frac{1+\alpha }{1-\alpha }}\right) +\frac{\alpha -1}{\alpha }v\left( 1+\frac{\alpha -1}{\alpha }v\right) ^{\frac{2\alpha }{1-\alpha }},
\end{equation}
where $v$ is the parameter given by \eqref{hubblespeed}. Moreover, from \eqref{eq:vspecpsf} and \eqref{eq:vastpsf} we have,
\begin{equation*}
\lim_{v\rightarrow v_{\ell \mathrm{max}}}\|v_{\mathrm{spec}}\|=1
\end{equation*}
and
\begin{equation}
\label{limvast}
\lim_{v\rightarrow v_{\ell \mathrm{max}}}\|v_{\mathrm{ast}}\|=\frac{1}{1+\alpha},
\end{equation}
where $v_{\ell \mathrm{max}}=\frac{\alpha}{1-\alpha}$ if $0<\alpha<1$, and $v_{\ell \mathrm{max}}=+\infty$ if $\alpha>1$.
\end{proposition}

It follows from \eqref{limvast} that the limiting astrometric speeds do not exceed $1$.
Spectroscopic and astrometric velocities are also represented in Figure \ref{alphagt1}.

\begin{remark}
For ``retarded comparisons'' (described in Section \ref{notation}), the kinematic and Fermi relative velocities of $u'_{\ell }$ must be calculated relative to $u^*=\left. \frac{\partial }{\partial t}\right| _{p^*}=(1,0)$, i.e., the 4-velocity of $\beta $ at $p^*=(\tau ^*,0)$. By \eqref{eqt1}, $\tau ^*=\tau^*\left( t_{\ell }(\tau ,\chi _{\ell }),\chi _{\ell }\right) $ is defined implicitly from
\begin{equation}
\label{eqtau*}
\int _{t_{\ell }(\tau ,\chi _{\ell })}^{\tau ^*} \frac{a(\tau ^*)}{a(t)}\frac{1}{\sqrt{a^2(\tau ^*)-a^2(t)}}\,\mathrm{d}t =\chi _{\ell },
\end{equation}
where $t_{\ell }(\tau ,\chi _{\ell })$ is given implicitly by \eqref{eqt1b}.
\end{remark}

\section{Concluding remarks}
\label{conclusion}

We have found general expressions for the Fermi, kinematic, astrometric, and spectroscopic  velocities of test particles experiencing radial motion relative to an observer  comoving with the Hubble flow (called the central observer) in any expanding Robertson-Walker cosmology.  Specific numerical calculations and formulas were given for cosmologies for which the scale factor, $a(t)=t^{\alpha}$, including both inflationary ($\alpha>1$) and non inflationary universes ($0<\alpha\leq1$). These include the radiation-dominated and matter-dominated universes, and models for dark energy (see \cite{power}).

It follows from Propositions \ref{findmetricA} and \ref{findmetric} that in principle, knowledge of either pair of the relative velocities for moving test particles at each spacetime point uniquely determines the geometry of the two dimensional spacetime via \eqref{fermipolar} and \eqref{opticalmetric}, and therefore the scale factor $a(t)$. Since the affine distance (i.e., the optical coordinate $\delta $) can be measured by parallax, and the frequency ratio can be found by spectroscopic measurements, the astrometric and spectroscopic relative velocities can, in principle, be determined solely by physical measurements, and so, they could confirm or contradict assumptions about the value of $a(t)$ for the actual universe.

Of the four relative velocities, only the Fermi relative velocity of a radially receding test particle can exceed the local speed of light of the observer (i.e., be superluminal), and this is possible at a spacetime point $(\tau, \rho)$, in Fermi coordinates, if and only if $-g_{\tau\tau}(\tau, \rho)>1$.

Under general conditions, the Hubble velocity of comoving test particles also become superluminal at large values of the radial parameter, $\chi$, and this is taken as a criterion for the expansion of space in cosmological models, and for the actual universe. By way of comparison, the Fermi relative velocity has both advantages and disadvantages to the Hubble velocity.  For comoving particles, both velocities measure the rate of change of proper distance away from the observer with respect to the proper time of the observer.  But for the Fermi velocity, the proper distance is measured along spacelike geodesics, while for the Hubble velocity the proper distance is measured along non geodesic paths.  In this respect the Fermi velocity is more natural and more closely tied to the observer's natural frame of reference, i.e., to Fermi coordinates in which locally the metric is Minkowskian to first order in the coordinates.  In addition, the notion of Fermi relative velocity, along with the other three relative velocities discussed in this work, are geometric and may be calculated in any spacetime, while the Hubble velocity is specific to Robertson-Walker cosmologies. The example of the Milne universe, discussed in the introduction, illustrates the limitations of the use of superluminal Hubble velocities as  indicators of expansion of space.

Cosmological models with a scale factor of the form $a(t)=t^{\alpha}$ for $ \alpha>0$, provide test cases for the use of the Fermi relative velocities of comoving test particles for understanding of expansion of space in general, and the effect of inflation and event horizons, in particular. Fermi coordinates are global in the non inflationary case, i.e., for $0< \alpha\leq1$ (see \cite{Klein11}), and maximal Fermi charts were shown in Section \ref{fermi} to extend up to (but not include) the cosmological event horizon, for the inflationary case with $\alpha>1$. Perhaps surprisingly, superluminal relative Fermi velocities of comoving particles exist only for the non inflationary cases, $0< \alpha<1$. Although not discussed in this work, the situation for the de Sitter universe is analogous.  There, the Fermi chart is valid only up to the cosmological horizon (see \cite{Klein11,CM,KC3}), and Fermi relative velocities of comoving test particles are necessarily subluminal.  One might expect that accelerating universes (in the sense that $\ddot{a}>0$) would allow for greater relative velocities, rather than impose lower speed limits.

Some insight into this phenomenon comes from the geometry of the simultaneous space slices, $\{\mathcal{M}_{\tau}\}$, and in particular, the dependence  of the proper radius, $\rho_{\mathcal{M}_{\tau}}$, on $\alpha$.  It was shown in \cite{Klein11} that for $0< \alpha\leq1$, superluminal relative Fermi velocities exist because ``there is enough space'' in the sense that the proper radius $\rho_{\mathcal{M}_{\tau}}$ of $\mathcal{M}_{\tau}$ satisfies the condition $\rho_{\mathcal{M}_{\tau}}>\tau$.  For $0< \alpha\leq1$, Fermi speeds increase with proper distance from the central observer, reaching
their limiting value $\rho_{\mathcal{M}_{\tau}}/\tau$ asymptotically.

It follows from \eqref{Mradius} that $\rho_{\mathcal{M}_{\tau}}(\alpha)$ decreases monotonically to zero as $\alpha\to +\infty$\footnote{The Hubble radius and proper distance to the event horizon in curvature coordinates also decrease monotonically with $\alpha$.}. This means that the spacelike geodesics orthogonal to the central observer's worldline reach the big bang, at $t=0$, in a proper distance that decreases with $\alpha$.  Since $\rho_{\mathcal{M}_{\tau}}/\tau<1$ for $\alpha >1$, one might expect, on this basis, the disappearance of superluminal Fermi relative velocities in the inflationary case.  We note, however, that in the inflationary cases, maximal Fermi speeds are achieved at proper distances less than $\rho_{\mathcal{M}_{\tau}}$.

Does space expand in the Robertson-Walker cosmologies with power law scale factors?  An affirmative answer may be given in the following sense. Let $\alpha$ be fixed; for any comoving geodesic observer, the Fermi space slices of $\tau$-simultaneous events, $\{\mathcal{M}_{\tau}\}$, that foliate spacetime up to the event horizon (for $\alpha>1$), or the entire space-time (for $0<\alpha\leq1$), have finite proper radii, $\rho_{\mathcal{M}_{\tau}}$,  that increase with the observer's proper time.  \\

\end{document}